\documentclass[aps,twocolumn,superscriptaddress,showpacs]{revtex4-1}
\usepackage{amsmath,amsfonts,amsbsy,amssymb,amsthm}
\usepackage[pdftex]{graphicx}
\usepackage{epstopdf}
\usepackage{array}

\usepackage[usenames,dvipsnames]{color}

\newcommand{\comment}[1]{}
\newcommand{\tr}{{\rm Tr}}
\newcommand{\ee}{{\rm e}}
\newcommand{\ii}{{\rm i}}

\newcommand{\ket}[1]{| #1 \rangle}
\newcommand{\bra}[1]{\langle #1 |}

\newcommand{\argmax}{\operatornamewithlimits{argmax.}}

\newcommand{\beq}{\begin{equation}}
\newcommand{\eeq}{\end{equation}}
\newcommand{\bea}{\begin{eqnarray}}
\newcommand{\eea}{\end{eqnarray}}

\newcommand{\rmd}{{\text d}}

\begin{document}

\title{Fidelity-optimized quantum state estimation}
\author{Amir Kalev}
\email{amirk@unm.edu}
\affiliation{Center for Quantum Information and Control, University of New Mexico, Albuquerque, NM 87131-0001, USA}
\author{Itay Hen}
\affiliation{Information Sciences Institute, University of Southern California, Marina del Rey, California 90292, USA}
\affiliation{Center for Quantum Information Science \& Technology, University of Southern California, Los Angeles, California 90089, USA}

\date{\today}

\begin{abstract}
We describe an optimized, self-correcting procedure for the Bayesian inference of pure quantum states.  By analyzing the history of measurement outcomes at each step, the procedure returns the most likely pure state, as well as the optimal basis for the measurement that is to follow. The latter is chosen to maximize, on average, the fidelity of the most likely state after the measurement. We also consider a practical variant of this protocol, where the available measurement bases are restricted to certain limited  sets of bases.  We demonstrate the success of our method by considering in detail the single-qubit and two-qubit cases, and  comparing the performance of our method against other existing methods.
\end{abstract}

\maketitle
\section{Introduction}\label{sec:intro}
Quantum technology is a fast developing area of both experimental and theoretical research. It has well-known applications in computation protocols~\cite{deutsch92,simon94,shor94,finilla94,grover96,farhi01} and communication protocols~\cite{bennett84,bennett92,bennett94,bennett96,lloyd97,bennett02}, as well as in the study of fundamental physical phenomena~\cite{lloyd96,kim10,islam11,barreiro11}.  A particular  interest in advanced technology is  that of devising protocols for basic primitives which are meant to
serve as building blocks for other, more involved applications. 
One example of this is quantum state estimation (QSE), whose purpose is to estimate the state of a quantum system based on a measurement record of a finite ensemble of identical systems.  

It is a common practice in QSE (and in many other protocols)  to have the quantum device complete a program which is then followed by a  classical processing step during which conclusions are drawn. However, with advanced technologies and the growing control over quantum devices, 
comes the obvious need (and possibility) to integrate these two steps into one indivisible 
unit within which the program and the data-processing steps are repeated successively -- feed-backing each other to form one efficient,
 self-correcting, self-executing quantum protocol. 
Indeed,  there has been a growing interest  in devising and implementing {\it in situ} protocols~\cite{jones91,jones94,fischer00,hannemann02,huszar12,tanaka12,mahler13,ferrie14,okamoto12,kravtsov13,wiebe13}, 
in particular using a Bayesian statistical inference, with various degrees of automation where the measurement at later steps depends on the information acquired in previous steps.  Some of these ideas were successfully implemented using current 
technology~\cite{mahler13} and were shown to improve accuracy quadratically over common  protocols. 

In any adaptive QSE protocol, the sequence of measurements depends on a choice of a  figure of merit.  For example, Husz{\'a}r and Houlsby~\cite{huszar12} considered an adaptive protocol using a Bayesian learning strategy derived from maximization of the average information gain. While such a strategy guarantees that each measurement yields, on average, the maximal information gain, it is not clear how it performs in terms of average infidelity (or equivalently, fidelity). Infidelity is a particularly important figure of merit in QSE as it quantifies the QSE inaccuracy (see, e.g., Ref.~\cite{mahler13} and references therein).

With the goal of minimizing estimation inaccuracy in mind, this paper discusses a QSE protocol based on Bayesian statistical inference whose objective is the maximization of the average fidelity.
The physical setup that we consider is one in which an `oven' prepares and emits copies of an unknown quantum pure state  of a given dimension $d$. We are then given the task of providing an estimation with the highest average fidelity to the state of the system by performing orthogonal projective (von Neumann) 
single-copy measurements.

The heart of the protocol consists of two subroutines: (1) the computation of the most likely candidate for the state emitted by the oven, and (2) based on the fidelity as a figure of merit, a computation of an optimal basis for the next measurement.  The resultant proposed protocol is therefore an optimized, fully automated, self-correcting algorithm. 

In the following, we describe the protocol as it applies to systems of finite-dimensional Hilbert spaces. We then study its performance by considering 
different situations and practical examples in order to elucidate the rather intuitive guiding principles of the protocol. In particular, we examine a `real-world' variant of the protocol, where the experiment is limited to measurements in certain restricted settings. We conclude with a summary and suggestions for possible follow-up research directions. 

\section{The protocol}\label{sec:protocol}
The main idea behind the protocol suggested here is the analysis, after each measurement, of the entire history of measurement outcomes thus far. The analysis yields the most likely state emitted by the oven and a basis for the next measurement. The latter is chosen such that measurement outcomes lead to new estimations whose average fidelity to the emitted state is maximal. Based on this analysis, a measurement is then performed in the calculated basis and the routine is repeated. 
By optimizing the measurement to be performed such that it will yield, at each step, the highest average fidelity, the proposed protocol maintains minimal inaccuracy in state estimation throughout its course.
 
\begin{figure}[t]
\begin{center}
\includegraphics[clip,width=1\linewidth]{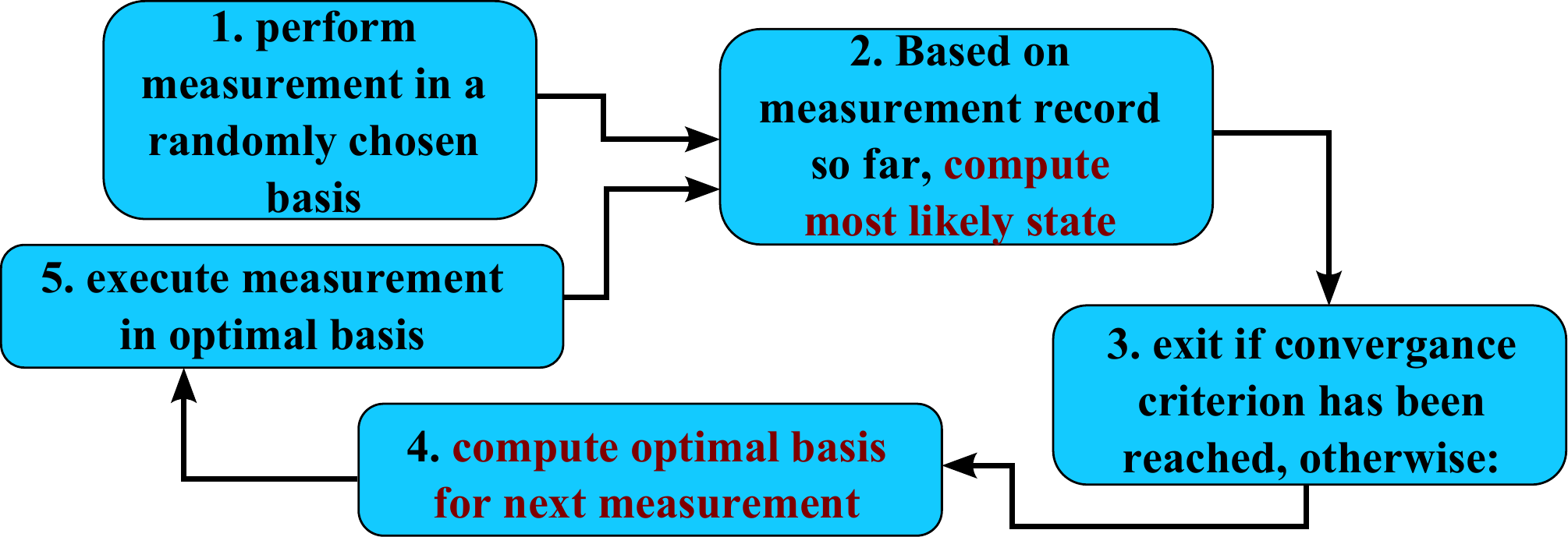}
\caption{{\bf Schematics of the optimized pure-state estimation protocol.}}
\label{fig:protocol}
\vspace{-0.7cm}
\end{center}
\end{figure}

We now define the steps of the protocol in detail. We assume, for simplicity, that no prior information concerning the distribution over which the oven is emitting the pure states is provided (the modifications required to address the more general case will be discussed below).  
The reader is referred to Fig.~\ref{fig:protocol} for a schematic diagram of the protocol. 
\begin{enumerate}
\item[Step 1:]
At first, no information about the state emitted by the oven has been acquired. 
The program thus executes a measurement in a randomly chosen basis, and records the result.
\item[Step 2:]
Based on all recorded outcomes so far, the program computes the (pure) state most 
likely to have yielded the sequence of measurement outcomes. 
The details of this computation are given below.  
We denote the most likely state after $k$ measurements by $|\Psi_k\rangle$.
\item[Step 3:]
If some predetermined stopping criterion has been reached, 
e.g., the fidelity $|\langle\Psi_{k-1}|\Psi_{k}\rangle|^2<1-\epsilon$, for a required accuracy level $\epsilon$, 
the program halts.    
\item[Step 4:]
After finding the most likely state, the program computes the optimal basis for the measurement that is to follow.  
The optimality condition is derived and discussed below. In practice, the experiment may be limited to a subset of measurement bases. 
We will also discuss the required adjustments to the protocol in the presence of such limitations. 
\item[Step 5:]
The program executes a measurement in the calculated basis, records the outcome, and returns to Step 2.
\end{enumerate}   

\subsection{Calculating the most likely pure state}

We now address the details of computing the pure state that represents our knowledge about the emitted state 
after $k$ iterations of the above protocol have taken place (and $k$ measurements have been executed). 

Let us consider the $k$-th iteration of the protocol. At each iteration, one measurement 
is performed on a single copy of the emitted state. Let \hbox{${\cal S}_k=\{|\phi_1\rangle, |\phi_2\rangle, \ldots, |\phi_k\rangle\}$} be the sequence of the $k$ measurement outcomes obtained and recorded thus far.  Based on the sequence ${\cal S}_k$, we are interested in finding the pure state that has the maximum fidelity with the emitted state. However, since we do not know what the emitted state is, we must average over all possible states, weighing each according to its probability of having been emitted (based on all recorded outcomes thus far).
The {\em most likely pure state of the system} $\ket{\Psi_k}$ would then be the pure state that has the maximal fidelity, on average, with all possible emitted states, namely:
\beq\label{eq:average fidelity}
|\Psi_k\rangle=\argmax_{\tilde{\psi}}\int\rmd P({\psi}|{\cal S}_k) \vert\langle{\tilde{\psi}}\vert{\psi}\rangle\vert^2 \,.
\eeq
Here, $\rmd{P}({\psi}|{\cal S}_k)$ is the (infinitesimal posterior) probability that the oven emitted a state $|{\psi}\rangle$ given the sequence of measurement outcomes obtained so far.

Interestingly, the above expression can be rewritten as
\beq\label{eq:average fidelity 2}
|\Psi_k\rangle=\argmax_{\tilde{\psi}}\bra{\tilde\psi}\rho_k\ket{\tilde\psi} \,,
\eeq
where $\rho_k$ is the state which best represents the totality of our knowledge so far about the system, given ${\cal S}_k$, 
\beq\label{eq:posterior state}
\rho_k=\int\rmd{P}({\psi}|{\cal S}_k) \ket{\psi}\!\bra{\psi} \,.
\eeq
Therefore, $|\Psi_k\rangle$ is the pure state with the highest fidelity to $\rho_k$, i.e., it is the state corresponding to the largest eigenvalue of $\rho_k$, 
\beq\label{eq:most likely pure}
|\Psi_k\rangle=\argmax_{\tilde{\psi}}\langle{\tilde{\psi}}|\rho_k|{\tilde{\psi}}\rangle=|\lambda^{\rm max}_{\rho_k}\rangle \,.
\eeq

To evaluate $\rmd{P}({\psi}|{\cal S}_k)$, we apply Bayes' formula  
\hbox{$\rmd{P}({\psi}|{\cal S}_k)={P}({\cal S}_k |{\psi}) \rmd{P}({\psi})/{P}({\cal S}_k)$}, where \hbox{${P}({\cal S}_k | {\psi})$} is the probability of obtaining the outcome sequence ${\cal S}_k$  provided that the emitted state is $\ket{\psi}$, and is given through Born's rule \hbox{${P}({\cal S}_k |{\psi})=\prod_{m=1}^k |\langle \phi_m |\psi \rangle|^2$}. 

The probability function $\rmd{P}({\psi})$ reflects our knowledge about the distribution over which the oven is choosing the states to emit, $\rmd {P}({\psi})=p({\psi})\rmd{\psi}$, where $p({\psi})$ is the corresponding probability density, and $\rmd\psi$ is a Haar-measure over the pure-state manifold.  
Consequently, the probability of obtaining the sequence of measurement outcomes is \hbox{${P}({\cal S}_k )=\int\rmd\psi{P}({\cal S}_k | {\psi})p({\psi})$}.
Combining the above, we arrive at: 
\beq \label{eq:dP}
\rmd P(\psi|{\cal S}_k) =\frac{\rmd\psi\; p(\psi)  \prod_{m=1}^k |\langle \phi_m |\psi \rangle|^2 }{\int\rmd\psi\; p(\psi)\prod_{m=1}^k |\langle \phi_m | \psi \rangle|^2  }\,.
\eeq
Now, since we are assuming that no prior information about the distribution over which the oven is emitting the states is given, we can set $p({\psi})=$ constant~\cite{footnote1}, which in turn allows us to identify the conditional (or posterior) 
infinitesimal probability  $\rmd P(\psi |{\cal S}_k)$ as 
\beq\label{eq:p(psi | s)}
\rmd P(\psi |{\cal S}_k) =\frac{\rmd \mathcal{P}_k}{\int \rmd \mathcal{P}_k} \,,
\eeq
where we have defined:
\beq \label{eq:Pk}
\rmd \mathcal{P}_{k} = \rmd{\psi} \prod_{m=1}^k |\langle \phi_m | {\psi} \rangle|^2\,.
\eeq
With the above definitions, we can rewrite the state that represents our knowledge about the system, Eq.~\eqref{eq:posterior state}, in the following way,
\beq\label{eq:posterior state 2}
\rho_k=\frac{\int\rmd \mathcal{P}_k\ket{\psi}\!\bra{\psi}}{\int \rmd \mathcal{P}_k}=\frac{\int\rmd{\psi} \prod_{m=1}^k |\langle \phi_m | {\psi} \rangle|^2\ket{\psi}\!\bra{\psi}}{\int \rmd{\psi} \prod_{m=1}^k |\langle \phi_m | {\psi} \rangle|^2} \,. 
\eeq

The above calculation can be generalized to include the case where some {\em a priori} knowledge about the distribution over which the oven emits the states is given, in which case $p({\psi})$ would no longer be constant. This may be of interest -- for example, when one wishes to certify that a particular target state has been experimentally realized. Additionally, the calculation of $\rho_k$ may be used to establish a stopping criterion for the protocol, e.g., by setting a threshold for the purity 
of the state, namely, $\tr[\rho_k^2]>1-\epsilon$ for a small predetermined $\epsilon$.

\subsection{Optimizing the basis of measurement}
Next, we describe a procedure to optimize the basis of the measurement to follow. 
The next measurement basis is calculated based on all the previous measurement outcomes. It is chosen to maximize the fidelity between our knowledge about the emitted state and the most likely state, as it would be computed after the measurement has been performed. 

Let ${\cal S}_{k-1}$ be the set of the ${k-1}$ measurement outcomes obtained and recorded, and let \hbox{${\cal \widetilde{B}}_k=\{ |\tilde{e}_{k,1}\rangle,|\tilde{e}_{k,2}\rangle,\ldots,|\tilde{e}_{k,d}\rangle \}$} be the orthonormal basis states of the $k$-th measurement over which the optimization would be carried out. 
Suppose now that a measurement has taken place in the ${\cal \widetilde{B}}_k$ basis with the outcome $|\tilde{e}_{k,n}\rangle$ (for some $1\leq n \leq d$).
The fidelity between the most likely state and our knowledge about the state of the system would be 
\beq \label{eq:rhokn}
\lambda_{\rho_{k,n}}^{\max} =\max_{\tilde{\psi}} \langle \tilde{\psi} | \left[
 \frac{\int \rmd \mathcal{P}_{k,n} |\psi\rangle \langle \psi |}{\int \rmd \mathcal{P}_{k,n}} \right]
 |\tilde{\psi}\rangle\,,
\eeq
 with
 $\mathcal{P}_{k,n}$ as defined in Eq.~\eqref{eq:Pk} with \hbox{$\mathcal{S}_{k,n}=\mathcal{S}_{k-1} \cup \{ |\tilde{e}_{k,n}\rangle\}$}. 
 
However, since we do not know which of the measurement outcomes will be realized, in order to optimize the basis ${\cal \widetilde{B}}_k$, we must consider all 
possible outcomes and weigh them according to their {\em a priori} probability  of occurring. 
This results in a weight-averaged maximal fidelity, $\sum_{n=1}^{d} P (\tilde{e}_{k,n} \vert \mathcal{S}_{k-1}) \lambda^{\max}_{\rho_{k,n}}$, where $P (\tilde{e}_{k,n}\vert \mathcal{S}_{k-1})$ 
is the probability of obtaining the outcome $|\tilde{e}_{k,n}\rangle$ given our knowledge so far:
\bea\label{eq:outcome prob}
P(\tilde{e}_{k,n}\vert \mathcal{S}_{k-1})&=&\bra{ \tilde{e}_{k,n}} \rho_{k-1}\ket{ \tilde{e}_{k,n}}  \\\nonumber
&=&\frac{\int \rmd \mathcal{P}_{k-1} |\langle \tilde{e}_{k,n} \vert \tilde{\psi}\rangle|^2}{\int \rmd \mathcal{P}_{k-1}}
= \frac{\int \rmd \mathcal{P}_{k,n}}{\int \rmd \mathcal{P}_{k-1}}
\eea
We define the {\em optimal basis of measurement} as the one which maximizes this weight-averaged maximal fidelity. Combining the expression above with the expression for $\lambda^{\max}_{\rho_{k,n}}$ in Eq.~(\ref{eq:rhokn}), the optimal basis of measurement simplifies to 
\begin{align}\label{eq:best basis}
&{\cal B}_k=
\argmax_{\{ |\tilde{e}_{k,1..d}\rangle\}}\sum_{n=1}^{d}\lambda^{\textrm{max}}_{\varrho_{k,n}}\,,
\end{align}
where $\varrho_{k,n}$ is the ``unnormalized'' density matrix
\beq
\varrho_{k,n} = \int \rmd \mathcal{P}_{k,n} |\psi\rangle \langle \psi | \,.
\eeq
The optimal basis thus neatly reduces to the basis which maximizes the sum of the largest eigenvalues of $\varrho_{k,n}$ with $n=1,2,\ldots,d$. 

By weighing equally the contributions from each of the possible measurement outcomes, Eq.~\eqref{eq:best basis} offers some intuition as to the optimal choice for the next basis of measurement:  That the probability to obtain any outcome, $\langle{\tilde{e}_{k,n}}|\rho_{k{-}1}|{\tilde{e}_{k,n}}\rangle$, is independent of the outcome label, i.e., $\langle{\tilde{e}_{k,n}}|\rho_{k{-}1}|{\tilde{e}_{k,n}}\rangle=1/d$, suggests that at each measurement step one should probe the system, in parameter space, in directions where our ignorance about the outcome is maximal, that is, in a basis that is {\it unbiased} to $\rho_{k{-}1}$.  
In the next section we study several explicit examples illustrating the above intuitive interpretation of this infidelity-minimizing cost function. 

\section{Optimized pure qubit state estimation}\label{sec:qubit}
As a case study, let us now consider the performance of the above protocol for the identification of a Haar-random pure qubit state.  Since for a qubit the Hilbert space is isomorphic to the Bloch sphere, the maximization at each stage can be performed with respect to the two real variables 
corresponding to the polar angle $\theta$ and the azimuthal angle $\varphi$ of the orientation of the Bloch vector of the qubit.

\subsection{First few measurements}
To gain some insight into the nature of the protocol, we examine in some detail its first few iterations as they apply to the qubit case. 
This will allow us to appreciate the intuitive interpretation of the protocol. 

Prior to the first measurement, no information about the state of the system is given. 
Thus, according to the proposed protocol, the first measurement is performed in a randomly chosen basis which we shall denote as 
\hbox{$\{|{\uparrow}\rangle,|{\downarrow}\rangle\}$}. For simplicity, let us denote the outcome direction of that first measurement by $|{\uparrow}\rangle$. 
At this point, the most likely pure state of the system is simply $|{\uparrow}\rangle$. 
(Hereafter, we represent density matrices and states in the  basis as \hbox{$\{|{\uparrow}\rangle,|{\downarrow}\rangle\}$}.)
To determine the basis in which the next ($k=2$) measurement will be performed, one should solve Eq.~\eqref{eq:best basis}, given the measurement record ${\cal S}_1=\{|{\uparrow}\rangle\}$. 
Upon doing so, the following set of degenerate solutions is found.
\beq
\{|e_2(\varphi)\rangle\}=\Bigl\{\frac{|{\uparrow}\rangle+\ee^{\ii\varphi}|{\downarrow}\rangle}{\sqrt 2},\frac{|{\uparrow}\rangle-\ee^{\ii\varphi}|{\downarrow}\rangle}{\sqrt 2}\Bigr\}.
\eeq
Without loss of generality, we take $\varphi=0$, so that the second measurement basis is 
\beq
\{|e_2\rangle\}=\Bigl\{|\pm\rangle=\frac1{\sqrt 2}(|{\uparrow}\rangle\pm|{\downarrow}\rangle)\Bigr\}.
\eeq
Performing the second measurement, we can again assume, without loss of generality, that the result $|+\rangle$ had been obtained. 
The most likely pure state describing the system, up to normalization, is
$|\Psi_2\rangle=\left( |\uparrow\rangle + |+\rangle \right)$, i.e., pointing in the direction $(\theta,\varphi)=(\pi/4,0)$ on the Bloch sphere.

To determine the basis of the next measurement, \hbox{$k=3$}, one should solve Eq.~\eqref{eq:best basis} given the outcome record ${\cal S}_2=\{|{\uparrow}\rangle,|{+}\rangle\}$.
In this case, we obtain the solution,
\beq
\{|e_3\rangle\}=\Bigl\{|{\pm\ii}\rangle=\frac1{\sqrt 2}(|{\uparrow}\rangle\pm\ii|{\downarrow}\rangle)\Bigr\}.
\eeq
As discussed in the previous section, we find that the optimal measurement basis at each step so far is in a direction that is unbiased to the  most likely state, namely 
$|\langle \Psi_{k-1} | e_{k,n}\rangle|^2=1/d=1/2$. 

At this point we can assume again that the third measurement result is $|{+\ii}\rangle$; this yields, for the most likely pure state describing the system, up to normalization, the state 
$|\Psi_3\rangle= \left( |\uparrow\rangle + |+\rangle + |{+\ii}\rangle \right)$ \textbf{--} i.e., 
pointing in the direction $(\theta,\varphi)=(\arccos\frac{1}{\sqrt{3}},\frac{\pi}{4})$ on the Bloch sphere.

So far, the optimal sequence of measurement bases has formed a set of mutually unbiased bases (MUBs). 
It follows then that the initial optimal probing sequence of the state naturally generates  mutually orthogonal directions on the Bloch sphere.
The fourth measurement basis is obtained by solving Eq.~\eqref{eq:best basis} given the measurement record \hbox{${\cal S}_3=\{|{\uparrow}\rangle,|{+}\rangle,|{+\ii}\rangle\}$}  which obviously cannot yield a fourth MUB.  Nonetheless, the solution in this case is found to be degenerate, corresponding to all directions that are unbiased to $|\Psi_3\rangle$, namely, \hbox{$|\langle \Psi_3 | e_{4,n}\rangle|^2=1/2$}  (equivalently, all vectors in a Bloch sphere orthogonal to the Bloch vector of $|\Psi_3\rangle$).  A summary of the first few measurements is given is Table~\ref{tbl:summary}.

\begin{table*}[t]
 \centering
\begin{tabular}{|c|| c | c |  c| c| m{2cm}|}
\hline
measurement & (unnormalized) most & most likely & next basis of & measurement &Bloch sphere \\
index $k$ & likely state $|\Psi_k\rangle$&  state fidelity & measurement $\{|e_k\rangle \}$& outcome &representation\\
\hline 
0 & --- & 1/2 & $\{|\uparrow, |\downarrow\rangle\}$ & $|\uparrow\rangle$ & \vspace{0.1cm}\begin{minipage}{0.1\textwidth}\includegraphics[width=0.95cm]{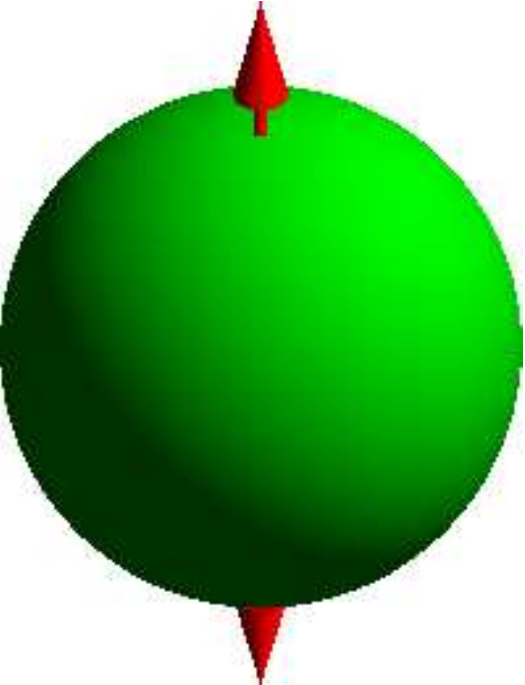}   \end{minipage} \vspace{0.1cm} \\
1 & $|\uparrow\rangle$ & 2/3 & $\{|+\rangle,|-\rangle\}$ &$|+\rangle$ & \begin{minipage}{0.1\textwidth}\includegraphics[width=1cm]{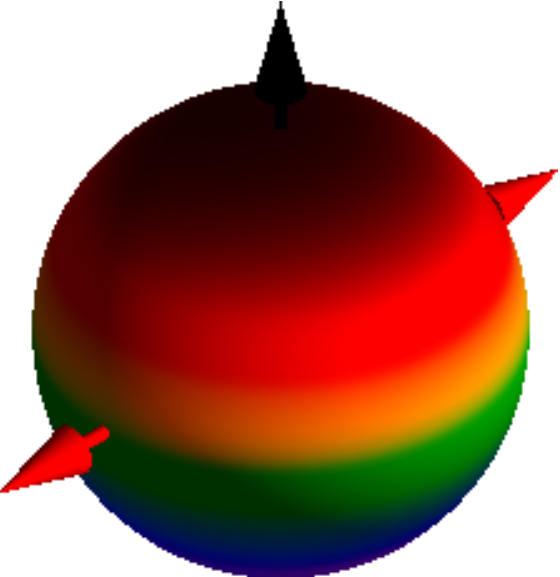}   \end{minipage} \vspace{0.1cm} \\
2 &  $\left( |\uparrow\rangle+ |+\rangle\right)$& $\frac{1}{2}+\frac{\sqrt{2}}{6}$& $\{|+\rm i\rangle,|- \rm i\rangle\}$ & $|+ \rm i\rangle$& \begin{minipage}{0.1\textwidth}\includegraphics[width=1cm]{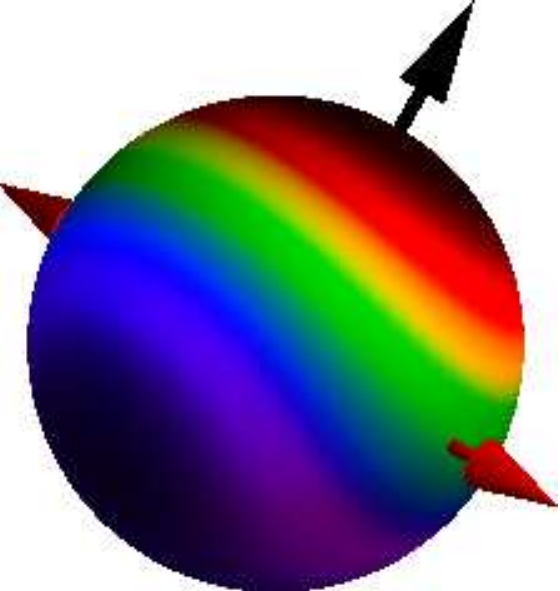}   \end{minipage} \vspace{0.1cm}\\
3 &  $\left( |\uparrow\rangle +|+\rangle+ |+\rm i\rangle \right)$& $\frac{1}{2}+\frac{\sqrt{3}}{6}$ &unbiased to $|\Psi_3\rangle$ & $~$& \begin{minipage}{0.1\textwidth}\includegraphics[width=.98cm]{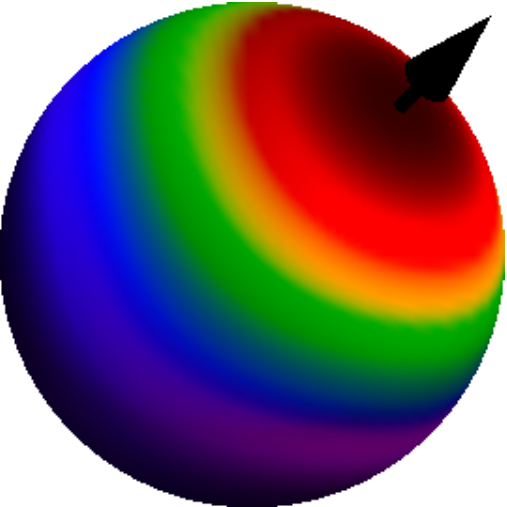}\end{minipage} \vspace{0.1cm}\\
\hline
\hline
\end{tabular}
\caption{(Color online) {\bf First few measurements of the single-qubit state estimation protocol.} The figures in the right-most column represent the posterior distribution given the measurement record. The color code red-to-blue indicates high-to-low probability regions. The black arrow represent the most likely state on the Bloch sphere, while the red arrows represent the fidelity-maximizing basis for the next measurement.}\label{tbl:summary}
\end{table*}

\subsection{Comparison with information-gain strategy}
As another illustrative example, we compare the performance of the proposed protocol against the performance of another strategy in which information-gain is used as a figure of merit (the reader is referred to Ref.~\cite{huszar12} for more details). 
Specifically, we analyze the first few steps of the two protocols for the case where the emitted qubit state is known {\em a priori} to lie in the $x{-}y$ plane of the Bloch sphere, i.e., that $p(\psi)$ is not a constant, but obeys $p(\psi)\sim \delta (\theta-\pi/2)$.

In the first step of the two protocols, no prior information about the direction of the qubit in the $x{-}y$ plane of the Bloch sphere is given. Therefore, without loss of generality, we assume that the first measurement is performed in the $\{|e_1\rangle\}=\{|{\pm}\rangle\}$ basis, and its outcome is $|{+}\rangle$.
As for the second measurement basis, it is interesting that both protocols yield the same next-measurement basis \textbf{--} namely, $\{|e_2\rangle\}=\{|{\pm\ii}\rangle\}$. Assuming, without loss of generality, that the outcome here is $|{+\ii}\rangle$, i.e., that \hbox{${\cal S}_2=\{|{+}\rangle,|{+\ii}\rangle\}$}, we move on to calculate the next basis of measurement. Here again, in both protocols, regardless of whether we optimize with respect to information-gain or infidelity, the same measurement direction is obtained for the third measurement, namely, \hbox{$\{|e_3\rangle\}=\{\frac1{\sqrt 2}(|{\uparrow}\rangle\pm{\rm e}^{\frac{3\ii\pi}{4}}|{\downarrow}\rangle)\}$}, which unsurprisingly is found to be unbiased to the most likely state \hbox{$|\Psi_2\rangle =\frac1{\sqrt 2}(|{\uparrow}\rangle\pm{\rm e}^{\frac{\ii\pi}{4}}|{\downarrow}\rangle)$}. 

It is in the fourth iteration, in which the next basis of measurement is computed based on the outcome record \hbox{${\cal S}_3=\{|{+}\rangle,|{+\ii}\rangle,\frac1{\sqrt 2}(|{\uparrow}\rangle{+}{\rm e}^{\frac{3\ii\pi}{4}}|{\downarrow}\rangle)\}$}, that the information-gain figure of merit~\cite{huszar12} and the infidelity figure of merit begin yielding different measurement directions. Here, numerical methods must be employed, yielding the directions \hbox{$\phi_{\text{max-fid.}}=2.9113(2)$} and \hbox{$\phi_{\text{inf-gain.}}=2.9322(2)$} for the fidelity-maximizing and information-gain methods, respectively. 
Importantly, we note that in using the fidelity-optimized strategy, we ensure that the fourth measurement basis has a maximal average fidelity, illustrating that maximal information-gain does not necessarily correspond to minimal inaccuracy.

\subsection{Asymptotic behavior}
\begin{figure}[t]
\begin{center}
\includegraphics[scale=0.75]{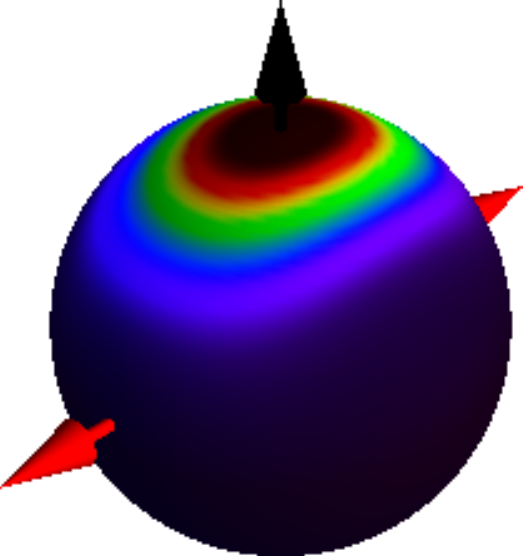}
\caption{(Color online) {\bf Measurement direction in the limit of many copies.} The figure presents the most likely vector (black arrow) and the optimal measurement basis (red arrows) for the sequence of measurements of Eq.~(\ref{eq:sk}) with $k_{1}=6, k_{2}=3$ and $k_{3}=1$. The false color plot corresponds to the probability density of $\rmd P(\psi | \mathcal{S}_k)$ for the above sequence of outcomes. The example illustrates that the optimal basis of measurement is mutually unbiased to the most likely state, and moreover, points in the `direction of least information' (in this example, the $y$ axis).}
\label{fig:variance}
\vspace{-0.7cm}
\end{center}
\end{figure}

In the following, we consider the basis of measurement that follows the sequence of outcomes
\begin{align} \label{eq:sk}
\mathcal{S}_{k}=\{&\underbrace{|{\uparrow}\rangle,\ldots,|{\uparrow}\rangle}_{k_1\text{ times}}, \underbrace{|{+}\rangle,\ldots,|{+}\rangle}_{k_2\text{ times}}, \underbrace{|{-}\rangle,\ldots,|{-}\rangle}_{k_2\text{ times}},\nonumber\\& \underbrace{|{+\ii}\rangle,\ldots,|{+\ii}\rangle}_{k_3\text{ times}},\underbrace{|{-\ii}\rangle,\ldots,|{-\ii}\rangle}_{k_3\text{ times}}\}\,,
\end{align}
where  $k=k_{1}+2 k_{2}+2 k_{3}$ is the total number of measurements. 
Although somewhat contrived, the above sequence provides the key ingredients to understanding the asymptotic behavior of the protocol, within the limit of a large number of copies.
First, we note that for any value of $k_{i}>0$, $i=1,2,3$, the most likely state is $|{\uparrow}\rangle$.  Employing the fidelity-optimizing protocol described above, it is easy to show that the optimal basis of the next measurement is $\{|\pm\rangle\}$ if $k_{2} > k_{3}$ and
 $\{|\pm \ii\rangle\}$ if $k_{2} < k_{3}$. (If $k_{2} = k_{3}$, both axes are equally optimal.) That is, the measurement basis vectors point in the direction in which we are ``most ignorant.''
This is illustrated in Fig.~\ref{fig:variance} which shows the color-coded probability measure $\rmd P(\Psi | \mathcal{S}_k)$ along with the most likely vector (black arrow) and the next measurement basis (red arrows) in the case $k_{1}=6, k_{2}=3$ and $k_{3}=1$. This is in accord with the intuitive interpretation of Eq.~(\ref{eq:best basis}) which was discussed previously. 

In particular, the above holds true for large $k$, in which case the probability measure $\rmd P(\psi | \mathcal{S}_k)$ is sharply peaked around the most likely state, i.e., around $|{\uparrow}\rangle$. In this case, the information-gain maximization approach~\cite{jones94,huszar12} dictates that the next measurement should be performed in a basis where one of its vectors points in a direction very close to the most likely state, i.e., the next measurement basis is approximately $\{\ket{{\uparrow}},\ket{{\downarrow}}\}$.
This basis differs considerably from the measurement basis obtained by maximizing the fidelity obtained by Eq.~(\ref{eq:best basis}).

\subsection{Numerical experiments\label{sec:numExpQubit}}
To test the full capabilities of the strategy proposed here on a qubit system, we conducted $N_{\text{exp}}=5000$ independent numerical experiments that execute the protocol for a few dozen iterations. 
In each experiment, a Haar-random pure state \hbox{$|\Psi\rangle$} was generated to simulate states emitted by the oven, followed by the application of the protocol discussed in Sec.~\ref{sec:protocol}. Measurements (one per iteration) were simulated numerically as well, using the generation of random numbers to produce measurement outcomes 
with the appropriate probabilities, i.e., with probabilities $|\langle e_{k,n}|\Psi\rangle|^2$. Here, $k$ is the measurement (or iteration) index and $n$ labels the basis state. As prescribed by the protocol, the next basis of measurement at every iteration was determined by maximizing the cost function, Eq.~(\ref{eq:best basis}), over all possible measurement bases. In the qubit case, since a measurement basis is uniquely defined by a single vector on the Bloch sphere, the optimization here was carried out over the polar and azimuthal angles $\theta$ and $\phi$. By employing several optimization methods, including exhaustive search, we found conjugate gradient to be the fastest and to yield the most accurate results. 

\begin{figure}[b]
\begin{center}
\includegraphics[width=1.02\linewidth]{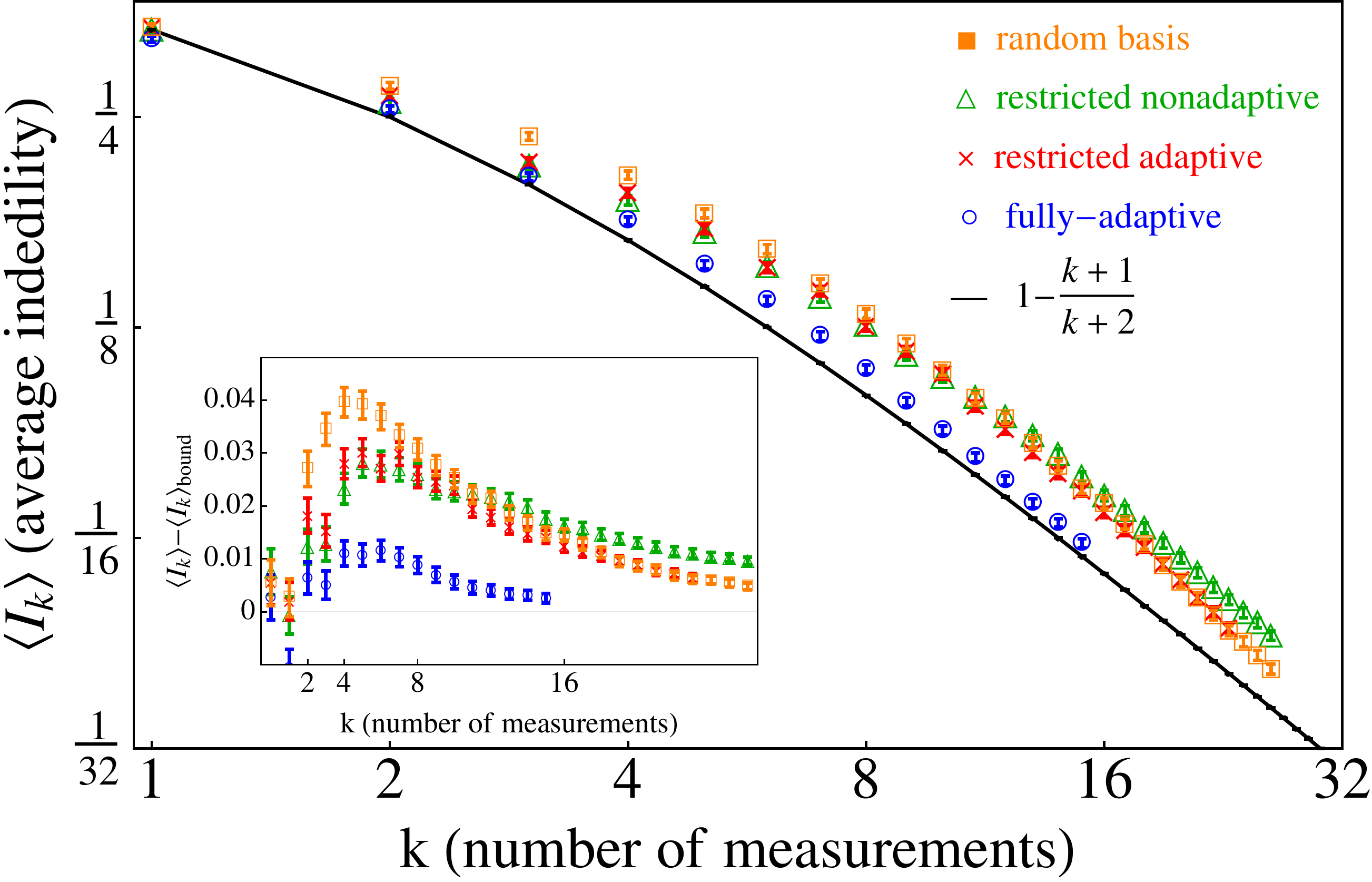}
\caption{(Color online) {\bf Average infidelity as a function of number of measurements for a single-qubit state estimation protocol.} Four strategies are compared: (1) the proposed protocol ({\color{blue}$\circ$}), (2)  a restricted-adaptive strategy ({\color{red}$\times$}), (3) a restricted-nonadaptive strategy with a fixed set of bases ({\color{ForestGreen}$\triangle$}) and (4) a random-basis strategy in which the measurement bases are chosen at random from the Haar-measure at each iteration ({\color{Orange}$\blacksquare$}). The bars indicate standard errors of the mean. The solid line is the theoretical bound achievable by an optimal collective measurement scheme on the entire ensemble of qubits.  
The inset shows the difference between the mean infidelity and the theoretical bound~\cite{massar95}, as a function of the number of measurements, for the various methods. The figure clearly illustrates the advantage that the proposed method has over others to quickly approach the theoretical bound.}
\label{fig:infid_1qubit}
\vspace{-0.7cm}
\end{center}
\end{figure}

We used the infidelity, 
\beq
I_k=1-|\langle \Psi | \Psi_k\rangle|^2,
\eeq
as our figure of merit to ascertain the success of our candidate state. 
For each numerical experiment, we recorded the infidelity $I$ as a function of number of iterations $k$. 
Finally, at every iteration $k$, we averaged the recorded infidelities over the $N_{\text{exp}}$ experiments, reporting $\langle I_k\rangle$. (We refer the reader
to Appendix~\ref{app:perm} for details concerning the actual calculation of the matrices $\varrho_{k,n}$.) The scaling of the average infidelity with the number of iterations, $k$, determines the performance of the method.  
The results of our experiments are summarized in Fig.~\ref{fig:infid_1qubit}.

Figure~\ref{fig:infid_1qubit}  shows how our protocol  ({\color{blue}$\circ$}) compares with three other strategies: a ``restricted-adaptive'' strategy ({\color{red}$\times$}), a ``restricted-nonadaptive'' strategy ({\color{ForestGreen}$\triangle$}), and a ``random-basis'' strategy where the measurement bases are chosen at random from the Haar-measure ({\color{Orange}$\blacksquare$}). In the  restricted-adaptive strategy the measurements are restricted to the set of Pauli bases, $\{|e_1\rangle\}=\{|{\uparrow}/{\downarrow}\rangle\}$, $\{|e_2\rangle\}=\{|\pm\rangle\}$, and $\{|e_3\rangle\}=\{|\pm{\rm i}\rangle\}$ and the optimization is carried out over this set of bases (see Appendix~\ref{app:restricted} for more details). 
In the restricted-nonadaptive strategy, the measurement directions, namely the Pauli $x, y$ and $z$ directions, are simply cycled through repeatedly.

As is clear from the figure, among the four protocols, the optimized protocol gives the best approximation by far to the theoretical bound.
This protocol is the solid line in Fig.~\ref{fig:infid_1qubit} and is only achievable by an optimal collective measurement scheme on an ensemble of $k$ qubits \hbox{$\langle I_k\rangle_{\text{bound}} = 1{-}\frac{k+1}{k+2}$}~\cite{massar95}. Additionally, as expected, we find  that the restricted-adaptive method performs better than the restricted-nonadaptive strategy and is somewhat comparable to the random basis strategy. The inset to Fig. 3 shows that the proposed protocol approaches the optimal bound faster than all the other simulated methods.

\section{Optimized two-qubit state estimation with local measurements}\label{sec:twoqubit}
To examine the performance of  our method in higher dimensions, we consider here the case of two-qubit state estimation. 
To grant our numerical experiment a more realistic flavor, we assume that the experimenter does not have access to all possible measurement settings, but rather to a small discrete subset of them. Specifically, we consider here a scenario where the experimenter is allowed to measure the qubits only locally in the Pauli bases. Thus, the measurement basis states of the qubits will be of the form $\{|e_n\rangle\}{\otimes}\{|e_m\rangle\}$, where each qubit set is one of the three bases $\{|{\uparrow}/{\downarrow}\rangle\}, \{ |\pm\rangle\}$ or $\{|\pm\ii\rangle\}$. 
In this case, the optimizing protocol, outlined in Sec.~\ref{sec:protocol}, can be executed as before, with the obvious modification that the `next basis of measurement' optimization is carried out only over the set of nine available bases. As before, we refer to this procedure as a restricted-adaptive procedure.

Similarly to the procedure described in Sec.~\ref{sec:numExpQubit}, we conducted \hbox{$N_{\text{exp}}=5000$} independent numerical experiments executing the protocol for about a dozen iterations.
In each experiment, a Haar-random two-qubit pure state \hbox{$|\Psi\rangle$} was generated to simulate states emitted by the oven~\cite{zyczkowski01}. The local measurements were simulated numerically as well to produce outcomes with the appropriate probabilities, i.e., with probabilities $|\langle e_{n,k}|\langle e_{m,k}|\Psi\rangle|^2$. Here, $k$ is the measurement (or iteration) index and $n,m$ label the basis states of each qubit.  

\begin{figure}[htp]
\begin{center}
\includegraphics[width=1\linewidth]{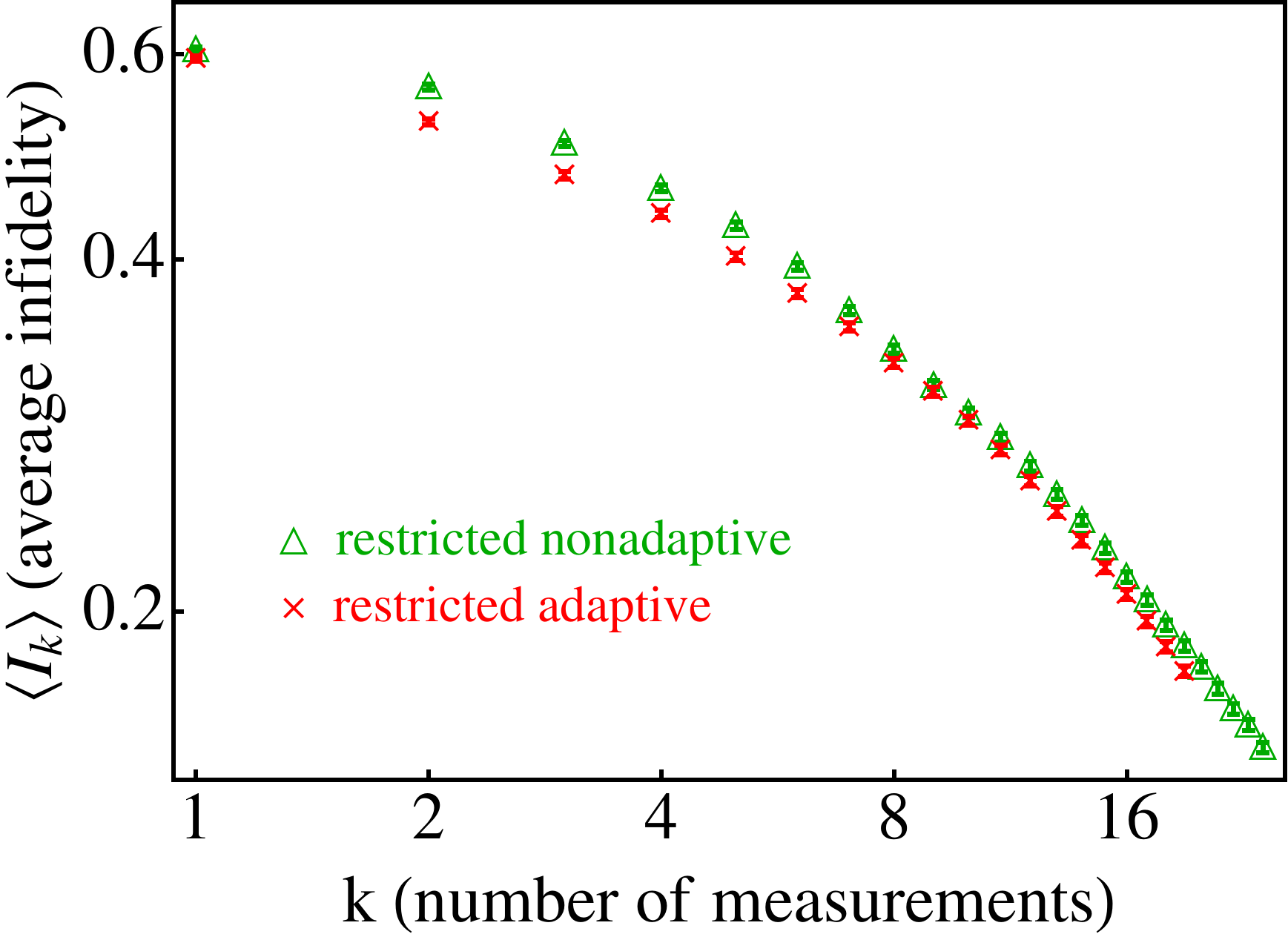}
\caption{(Color online) {\bf Average infidelity as a function of the number of measurements for a two-qubit state estimation protocol.} Two strategies are compared: (1) a restricted-adaptive strategy ({\color{red}$\times$}), and (2) a restricted-nonadaptive strategy ({\color{ForestGreen}$\triangle$}). The error bars are smaller than the symbols. Even on a restricted set of measurement bases, we find an appreciable advantage in using the adaptive strategy over the nonadaptive one.}
\label{fig:infid_2qubit}
\vspace{-0.4cm}
\end{center}
\end{figure}

Fig. ~\ref{fig:infid_2qubit} demonstrates how the restricted-adaptive protocol of local Pauli measurements  ({\color{red}$\times$}) compares with a nonadaptive strategy in which the available measurements bases are cycled through repeatedly ({\color{ForestGreen}$\triangle$}). As is clear from the figure, in this scenario too, there is an appreciable gain from the use of the adaptive optimizing protocol. 

\section{Conclusions and further remarks}\label{sec:conc}
We described a fidelity-maximizing, optimized, fully automated quantum state tomography procedure for pure states.  
The optimality of the protocol is achieved by requiring that each choice of measurement basis results in an estimation that has the highest fidelity to the state representing our knowledge about that emitted by the oven. 

We illustrated the power and intuitive interpretation of the proposed technique by considering several theoretical and practical scenarios. Since the protocol described here is optimized at every iteration, no post-processing of the data is required.  It would therefore be of particular interest to apply it as an {\it in situ} procedure in integrated devices for quantum tomography. Such a device would act like a `robot,' correcting itself as needed to find the state emitted by an oven, based on the information it is gathering.  In addition, since the protocol is optimal at every stage, it may be stopped at any stage, or after a chosen criterion of confidence or convergence is met. Therefore, no prior knowledge of the number of copies provided by the oven is necessary. In light of the above, we believe that our figure of merit, Eq.~(\ref{eq:best basis}), and its generalization to mixed states, could become a useful practical tool for experimentalists to achieve efficient state estimation.  

Moreover, we find no obvious reason that the above protocol and formalism could not be straightforwardly generalized to the setting of mixed states or general quantum measurements (POVMs), as well as to cases where prior information about the state-emitting oven is given. We shall consider these generalizations elsewhere.

Finally, we note that an interesting property of the strategy presented here is that it naturally ``generates'' in its first few iterations, MUBs as bases of measurements; unlike other methods, the protocol does not require {\em a priori} knowledge of MUBs. Based on this observation, we conjecture that for {\it any} $d$-dimensional system, the first few iterations of our protocol would optimally consist of measuring the system in a sequence of MUBs. For those Hilbert space dimensions for which the number of MUBs is unknown (i.e., if $d$ is not a power of prime~\cite{wootters89,tal02,klap04,durt05}), this property may be leveraged to answer several outstanding questions associated with the enumeration of MUBs. This avenue of research is currently being pursued by the authors. 

\begin{acknowledgments} 
We thank Gabriel Durkin, Christopher Ferrie and Nathan Wiebe for useful comments and discussions.  AK acknowledges support under NSF grant number PHY-1212445. IH acknowledges support under ARO grant number W911NF-12-1-0523. This research used resources of the Oak Ridge Leadership Computing Facility at Oak Ridge National Laboratory, which is supported by the Department of Energy's Office of Science under Contract DE-AC05-00OR22725.
\end{acknowledgments}

\appendix
\section{Calculation of $\varrho_{k,n}$}\label{app:perm}
Following Ref.~\cite{shchesnovich14}, we provide the necessary expressions required for calculating $\varrho_{k,n}$.
The basic identity we utilize is
\begin{align}
&\langle\phi_k|\cdots\langle\phi_1|\Bigl[\int\rmd\psi(|\psi\rangle\!\langle\psi|)^{\otimes k}\Bigr]|\phi_1\rangle\cdots|\phi_k\rangle\nonumber\\&=\frac{(d-1)!}{(k+d-1)!}{\rm Per}({\cal M}),
\end{align}
where $|\psi\rangle$ is an element in a $d$-dimensional Hilbert space and ${\rm Per}({\cal M})$ is the permanent of the $k\times{k}$ matrix ${\cal M}$ whose elements are ${\cal M}_{i,j}=\langle\phi_i|\phi_j\rangle$.
In our case 
\beq
\varrho_k=\langle\phi_k|\cdots\langle\phi_1|\Bigl[\int\rmd\psi(|\psi\rangle\!\langle\psi|)^{\otimes k+1}\Bigr]|\phi_1\rangle\cdots|\phi_k\rangle,
\eeq
and therefore the matrix elements of $\varrho_k$ in some basis $\{|z_i\rangle\}$ are given by \begin{align}
\varrho_{k,ij}&\equiv\langle z_i|\varrho_k|z_j\rangle\nonumber\\&=\langle z_i|\langle\phi_k|\cdots\langle\phi_1|\Bigl[\int\rmd\psi(|\psi\rangle\!\langle\psi|)^{\otimes k+1}\Bigr]|\phi_1\rangle\cdots|\phi_k\rangle|z_j\rangle\nonumber\\&=\frac{(d-1)!}{(k+d)!}{\rm Per}({\cal A}^{(ij)}),
\end{align}
where ${\cal A}^{(ij)}$ is a $k{+}1\times{k{+}1}$ matrix that can be written as follows,
\beq
{\cal A}^{(ij)} =\begin{pmatrix}
  {\cal M} & V^{(j)} \\
  (V^{(i)})^\intercal &\delta_{i,j} \end{pmatrix},
\eeq
and $(V^{(i)})^\intercal=(\langle e_i|\phi_1\rangle,\langle e_i|\phi_2\rangle,\ldots,\langle e_i|\phi_k\rangle)$ and similarly for  $V^{(j)}$. 

Suppose we obtained $k-1$ measurement outcomes. Then, the most likely pure state that describes the system is the state that corresponds to the maximal eigenvalue of $\varrho_{k{-}1}$. 

To optimize the next measurement basis we first need the eigenvalues of $\varrho_{k,n}$ as a function of the $k$th measurement outcome, $|\tilde{e}_{k,n}\rangle$, $n=1,\ldots,d$. The matrix elements of $\varrho_{k,n}$ are written (in some computational bases $\{|z_i\rangle\}$) as
\begin{align}
\varrho_{k,n,ij}&\equiv\langle z_i|\varrho_{k,n}|z_j\rangle\nonumber\\&=\langle z_i|\langle \tilde{e}_{k,n}|\langle\phi_{k{-}1}|\cdots\langle\phi_1|\nonumber\\&\times\Bigl[\int\rmd\psi(|\psi\rangle\!\langle\psi|)^{\otimes k+1}\Bigr]|\phi_1\rangle\cdots|\phi_{k{-}1}\rangle|\tilde{e}_{k,n}\rangle|z_j\rangle\nonumber\\&=\frac{(d-1)!}{(k+d)!}{\rm Per}({\cal A}^{(ij)})\,.
\end{align}

\section{Single-qubit case: restricted set of measurement bases}\label{app:restricted}
In the special case where the set of possible measurement directions is restricted to the $x, y$ and $z$ directions on the Bloch sphere, one does not need to resort to the costly permanents discussed above. In this scenario, calculations simplify. 
For example, the product $\prod_{i=1}^{k} |\langle \hat{n}(\theta_i,\phi_i)|\hat{n}(\theta,\phi)\rangle|^2$ takes the form:
\begin{align}
&\prod_{i=1}^{k} |\langle \hat{n}(\theta_i,\phi_i)|\hat{n}(\theta,\phi)\rangle|^2=\frac1{2^{m}}
\left(1+\cos \theta\right)^{m_1}
\left(1-\cos \theta\right)^{m_2}\nonumber\\&\times
\left(1+\sin \theta \cos \phi \right)^{m_3}
\left(1-\sin \theta \cos \phi \right)^{m_4}
\left(1+\sin \theta \sin \phi \right)^{m_5}\nonumber\\&\times
\left(1-\sin \theta \sin \phi \right)^{m_6} \,,
\end{align}
where the $m_i$ denote the number of times the outcomes $\ket{{\uparrow}}, \ket{{\downarrow}},\ket{{+}},\ket{{-}},\ket{{+\ii}}$, and $\ket{{-\ii}}$ have been obtained, respectively, and $m=\sum_{i=1}^{6} m_i$. The above product can be expanded to give:
\begin{align}\label{eq:sums}
&\prod_{i=1}^{k} |\langle \hat{n}(\theta_i,\phi_i)|\hat{n}(\theta,\phi)\rangle|^2\nonumber\\&=
\sum_{k_1,k_2,\ldots,k_6=0}^{m_1,m_2,\ldots,m_6} \prod_{i=1}^{6} {m_i\choose{k_i}} (-1)^{k_2+k_4+k_6} \left(\cos\theta\right)^{k_1+k_2}\nonumber\\&\times\left(\sin\theta\right)^{k_3+k_4+k_5+k_6}
\left(\cos\phi\right)^{k_4+k_5}\left(\sin\phi\right)^{k_5+k_6}.
\end{align}
In this form, all the integrals (for calculating $\varrho_k$ or $\int \rmd \mathcal{P}_k$) will be of the form
\beq
\int_0^{\pi} \rmd \theta \cos^m \theta\sin^n\theta = \frac{1 + (-1)^m}{2} \frac{\Gamma\left(\frac{1 + m}{2}\right) \Gamma \left( 1 + \frac{n-1}{2}\right)}{\Gamma\left(\frac{2 + m + n}{2}\right)}\,,
\eeq
and
\begin{align}
\int_0^{2\pi} \rmd \phi \cos^m \phi\sin^n\phi &= \left[1 + (-1)^n\right] \nonumber\\&\times\frac{1 + (-1)^m}{2} \frac{\Gamma\left(\frac{1 + m}{2}\right) \Gamma \left(\frac{n+1}{2}\right)}{\Gamma\left(\frac{2 + m + n}{2}\right)}\,.
\end{align}
Since the integrals are performed explicitly, expressions for obtaining the various matrix entries of $\varrho_k$ and $\int \rmd \mathcal{P}_k$ reduce in this case to the more-easily calculable multiple sums over the $k_i$'s
of Eq.~(\ref{eq:sums}).

\end{document}